\documentclass[twocolumn,aps,pra,groupedaddress,showpacs,floatfix]{revtex4}
\usepackage{epsfig,amssymb,amsmath}
\begin{document}

\newcommand{\ket}[1]{ | \, #1  \rangle}
\newcommand{\bra}[1]{ \langle #1 \,  |}
\newcommand{\eins}{\mbox{$1 \hspace{-1.0mm}  {\bf l}$}}
\newcommand{\proj}[1]{\ket{#1}\bra{#1}}
\providecommand{\ran}{\rangle}
\providecommand{\lan}{\langle}
\providecommand{\calh}{{\cal H}}
\providecommand{\cale}{{\cal E}}

\title{Quantum relative positioning in Hilbert space}

\author{Vittorio Giovannetti and Rosario Fazio}
\affiliation{NEST-INFM \& Scuola Normale Superiore,
	piazza dei Cavalieri 7, I-56126 Pisa, Italy.}

\begin{abstract}
A new class of state transformations that are quantum mechanically
prohibited is introduced. 
These can be seen as the generalization of the universal-NOT 
transformation which, for all pure inputs state of a given Hilbert space  
produces pure outputs whose projection on the original state 
is fixed to a value smaller than one. The case of not pure output
states is also addressed. We give an application of these 
transformations in the context of separability criteria.
\end{abstract}
\maketitle

Quantum mechanics imposes severe constraints on the way we can transform a 
physical system~\cite{WERNERBOOK}. Perhaps the most popular of these quantum no-go 
theorems is related with the impossibility of cloning~\cite{WOOTTERS} an unknown 
quantum state $|\psi\ran$. One cannot as well implement the Universal-Not (U-NOT)
operation 
which transforms $|\psi\ran$ into an orthonormal vector $|\psi_{\perp}\ran$~\cite{PASQ,BUZEK}.
Along the same line we ask if it is 
possible to find universal operations that map an unknown  state of a system S on a portion of 
the Hilbert space ${\cal H}_S$ of S such that some prescription is satisfied~\cite{NOTAQRP}.
We refer to this question as the {\em Quantum Relative Positioning} (QRP) with respect to an 
unknown initial state. 
There are several motivations for addressing it.
First of all the possibility to identify a new class of forbidden transformations 
which generalize the U-NOT operation allows us to
frame what is known for the U-NOT in a broader context~\cite{WERNERBOOK}. 
Moreover an analysis  of the impossible QRP transformations 
may be relevant in the study of the entanglement. 
By definition the QRP transformations are in fact positive 
(they map quantum states of S into quantum states) but in general they will not be 
completely positive. Thus one may try to employ QRP operations for characterizing the 
entanglement of composite systems as in Refs.~\cite{HORO3,HORO,BUZEK1}: an example
of this approach will be discussed at the end of the paper.

\begin{figure}[t]
\begin{center}
\epsfxsize=.9\hsize\leavevmode\epsffile{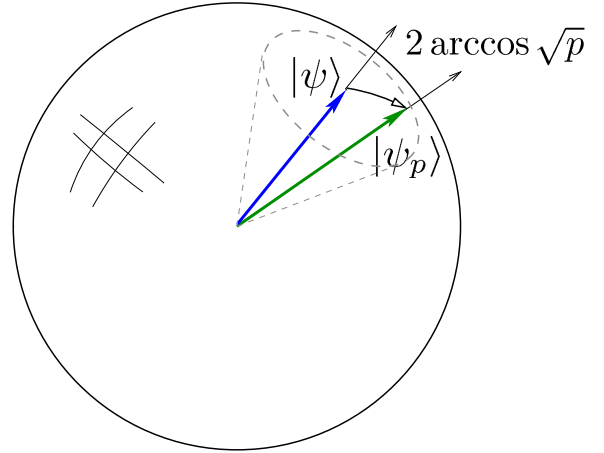}
\end{center}
\caption{Proper quantum mover
	$\mbox{PQM}(p)$  on a qubit. This transformation takes 
 	an unknown state $|\psi\ran$ of the qubit
	into a pure state $|\psi_p\ran$ whose projection on the 
	original vector is equal to $p$.
	Proper quantum movers should not be confused with ordinary rotations:
	the later are unitary transformations defined with respect to fixed  external 
	axis that do not depend on the input state $|\psi\ran$.
	If $p<1$ the  transformation $\mbox{PQM}(p)$
	cannot be physically implemented,  
	i.e. it cannot
	be obtained by means of unitary transformations acting on a (possible larger) 
	Hilbert space~\cite{CHUANG}. 
	For $p=0$, $\mbox{PQM}(0)$
	performs a universal-not which maps $|\psi\rangle$ to its antipode.}
\label{f:fig1}
\end{figure}

A complete characterization of the QRP transformations is quite a
 challenging task. In the following 
we will focus on an important class of these operations which contains  the U-NOT as 
a special case. These are the  {\em Generalized Quantum Movers} 
$\mbox{GQM}(p)$ which transform an unknown input pure state $|\psi\ran$ of S into some output 
density matrix  $\rho_p$ (possibly depending on $|\psi\rangle$) whose fidelity~\cite{JOZSA} 
with respect to $|\psi\ran$ is equal to a fixed value $p<1$, i.e.
\begin{equation}
	F(|\psi\ran, \rho_p) \equiv \bra{\psi} \rho_p |\psi\rangle =p \;,
\label{movers1}
\end{equation}
for all $|\psi\rangle$ of ${\cal H}_S$ (the case $p=1$ is trivially satisfied by the 
identity map). An important subset of the transformations $\mbox{GQM}(p)$
is the set of the {\em Proper Quantum Movers} $\mbox{PQM}(p)$ 
which satisfy Eq.~(\ref{movers1}) with pure outputs states $\rho_p=|\psi_p\rangle \langle \psi_p|$.
More precisely, a proper quantum mover $\mbox{PQM}(p)$ 
takes  $|\psi\rangle$ into a pure  vector $|\psi_{p}\ran$ (depending on $|\psi\rangle$)
whose  projection  $|\bra{\psi} \psi_p\ran |^2$ on $|\psi\ran$ is equal to $p$.
A simple geometrical representation of  $\mbox{PQM}(p)$ 
emerges in the case of qubits: here $|\psi_p\rangle$ is one of the points where the Bloch 
sphere intercepts the cone of angular opening  $2 \arccos{\sqrt{p}}$ originating from 
the center and oriented along $|\psi\rangle$ (see Fig.~\ref{f:fig1}). 

In the following we will prove the impossibility of building a machine 
which implements a proper quantum mover $\mbox{PQM}(p)$ on a physical system S 
by showing that this transformation 
does not correspond to any
linear, completely positive, trace preserving map~\cite{CHUANG} acting on the 
Hilbert space ${\cal H}_S$ of S.
Since $\mbox{PQM}(0)$ coincides with the U-NOT transformation,  for $p=0$
our results reproduce those of Refs.~\cite{PASQ,BUZEK}. However, the 
impossibility of implementing $\mbox{PQM}(p)$ for  $p\neq 0$ seems to be a novel 
quantum no-go theorem whose existence cannot be established from the impossibility of 
performing U-NOT (e.g. given the arbitrariness in selecting the output state 
$|\psi_p\rangle$, successive applications of  $\mbox{PQM}(p)$
on $|\psi\rangle$ do not necessarily produce a rotation of this state 
into an orthogonal vector). The scenario changes 
when the  hypothesis on the purity of the output state is relaxed. In fact, we will show that  
$\mbox{GQM}(p)$ with mixed outputs 
 can be physically implemented on S if and only if $p$ 
is greater than a threshold value $p_{\mbox{\small t}}(d)$ which depends upon 
the dimension $d$ of the Hilbert space ${\cal H}_S$. Moreover  a simple connection 
between the generalized quantum 
movers and the U-NOT will be established by showing 
that if the U-NOT were possible then also the $\mbox{GQM}(p)$ 
could be also realized. 

\section{Pure outputs}\label{s:uno}

Proving that proper quantum mover $\mbox{PQM}(p)$
cannot be implemented on a physical system S is equivalent to showing the following:

{\em  Given $p \in [0,1[$ there is no deterministic physical process 
$\cal E$ which transforms  an unknown pure vector $|\psi\ran$ of a generic 
Hilbert space ${\cal H}_S$ according to the mapping
\begin{eqnarray}
	|\psi\ran \;  \stackrel{\cal E}{\longrightarrow} \; |\psi_p\ran \label{THETRA} \;,
\end{eqnarray}
where $|\psi_p \ran$ is a pure state of ${\cal H}_S$ depending on $|\psi\rangle$ 
which satisfies the identity
\begin{eqnarray}
	|\bra{\psi} \psi_p\ran |^2=  p \;.
\label{unnot}
\end{eqnarray}}
The adjective ``deterministic'' in the hypothesis plays an important  role.
On one hand it allows us to exclude from the analysis those processes  which 
implement~(\ref{THETRA}) probabilistically, i.e. allowing some probability of failure.
As in the case of the cloning and the U-NOT machines~\cite{DEMA,PASQ,BUZEK,BUZEK1}
such transformations are not prohibited by quantum mechanics and the above no-go theorem 
does not apply to them. In addition the term ``deterministic'' indicates that here 
we are considering scenarios where the output state $|\psi_p\ran$ associated with 
$|\psi\ran$ is uniquely determined: i.e. if we run the transformation many times 
on various copies of a certain input $|\psi\ran$, the output state will be always 
the same vector $|\psi_p\ran$. This point is important since the vectors $|\psi_p\ran$ 
which satisfy the property~(\ref{unnot}) form a dense subset of ${\cal H}_S$. A transformation 
(\ref{THETRA}) which randomly picks up one of these $|\psi_p\ran$ is described by 
a $\mbox{GQM}(p)$ map whose output state is not pure: these operations will be discussed in the 
next section where we will see that the possibility of implementing them depends 
explicitly on the dimension of ${\cal H}_S$.

The most general deterministic physical transformations~(\ref{THETRA}) 
of the system S  are the linear, completely positive,
trace preserving (LCPT) maps $\cal E$ which for all $|\psi\ran \in {\cal H}_S$ 
give
\begin{eqnarray}
	{\cal E}(|\psi\rangle \langle \psi|) 
	&=&| \psi_p \rangle \langle \psi_p | 
\label{cpt}\;,
\end{eqnarray}
with $|\psi_p\ran$ unit vector of ${\cal H}_S$ such that Eq.~(\ref{unnot}) 
applies~\cite{NOTA0}.
In the qubit case ($d=2$)  and for $p<1/3$ 
the impossibility of implementing the transformation~(\ref{cpt}) 
can be established  by considering  the maximal fidelity achievable in performing 
an approximated U-NOT~\cite{BUZEK,PASQ}. In fact, by expressing the trace of 
${\cal E}(|\psi\rangle \langle \psi|)$ in the orthonormal basis $|\psi\rangle$, 
$|\psi_\perp\rangle$ of ${\cal H}_S$ we get 
\begin{eqnarray}
	\langle \psi_\perp | {\cal E}(|\psi\rangle
	\langle \psi|) |\psi_\perp\rangle = 1 - p\;.
\label{ULTIMA}
\end{eqnarray}
The thesis then follows by observing that the left-hand side term of this 
expression is the fidelity of the output state of $\cal E$ with respect to 
the orthogonal vector $|\psi_\perp\rangle$ which according to~\cite{BUZEK,PASQ} is
upper bounded by $2/3$. 

The above argument does not apply if $d > 2$ and for generic $p$. One can still 
prove that no  transformations~(\ref{cpt})  exist by considering a unitary 
representation~\cite{CHUANG} of $\cale$, i.e.
 \begin{eqnarray}
{\cal E}(| \psi\rangle \langle \psi|) = 
\mbox{Tr}_A \big[ \; U \; \left( 
\ket{\psi}\bra{\psi} \otimes \ket{0}_A\bra{0}\right) \; U^\dag \;\big]\;, \label{cpt1}
\end{eqnarray}
where $U$ is a unitary operator acting on S and on an ancillary system A,  
while $|0\ran_A$ is a fiduciary state of A. According to Eq.~(\ref{cpt}) the pure states 
of ${\cal H}_S$ are mapped into pure states of the same space, this implies that
the state $U \; \ket{\psi} \otimes \ket{0}_A$ cannot be entangled, i.e.
\begin{eqnarray}
U \; \ket{\psi} \otimes \ket{0}_A = \ket{\psi_p} \otimes \ket{\phi(\psi)}_A \label{cpt2}
\end{eqnarray}
with $\ket{\phi(\psi)}_A$ a pure state of A which, in principle, depends on the input 
state $|\psi\ran$ of S. By applying this relation to the unit vector
\begin{eqnarray}
|\psi\ran = \alpha \; |k\ran + \beta \; |k^\prime\ran\label{VETTORI}\;,
\end{eqnarray}
(with $|k\ran$, $|k^\prime\ran$ orthonormal states of ${\cal H}_S$ and $\alpha$, $\beta$ complex
such that $|\beta|^2=1-|\alpha|^2$) and using the linearity of $U$ one gets
\begin{eqnarray}
\ket{\psi_p} \otimes \ket{\phi(\psi)}_A = 
\alpha\; |k_p\ran \otimes \ket{\phi(k)}_A + \beta \; |k_p^\prime\ran \otimes
\ket{\phi(k^\prime)}_A \nonumber \\
\label{cpt3}
\end{eqnarray}
with $|k_p\ran$, $\ket{\phi(k)}_A$ and  $|k_p^\prime\ran$, $\ket{\phi(k^\prime)}_A$
defined as in Eq.~(\ref{cpt2}). In particular $|k_p\ran$ and $|k_p^\prime\ran$ are, 
respectively, the {\em unit vectors} of ${\cal H}_S$ which describes the 
outputs of the map $\cale$ associated with $\ket{k}$, $\ket{k^\prime}$: by  hypothesis 
they satisfy the relations $\bra{k_p}k\ran = \sqrt{p} e^{i \phi}$, 
$\bra{k^\prime_p}k^\prime \ran = \sqrt{p} e^{i \phi^\prime}$ for some 
phases $\phi$ and $\phi^\prime$.
By projecting the left hand side of Eq.~(\ref{cpt3}) on $|\psi\ran$ we derive the following 
identity
\begin{eqnarray}
p&=& \| \big( |\alpha|^2  \sqrt{p} \; e^{i \phi} +
 \alpha \beta^*\; \bra{k^\prime} k_p\ran  \; \big) |\phi(k)\ran_A \nonumber \\
&&+ \big(|\beta|^2   \sqrt{p} \; e^{i \phi^\prime}
+ \alpha^*\beta\;  \bra{k} k^\prime_p\ran \;  \big) |\phi(k^\prime)\ran_A \|^2 \;,\label{cpt10}
\end{eqnarray}
which must hold for all the complex amplitudes $\alpha$ and $\beta$.
One can verify that 
this condition is satisfied if and only if
\begin{eqnarray}
 \bra{k^\prime} k_p\ran  = \bra{k} k^\prime_p\ran = 0 
\label{cpt11}\;,
\end{eqnarray}
(see Appendix~\ref{appendixA} for details).
The impossibility of  the transformation $\cal E$
finally follows by observing that Eq.~(\ref{cpt11}) is in contradiction
with $|k_p\ran$ being a 
unit vector of ${\cal H}_S$. In fact since Eq.~(\ref{cpt11})
applies for all $|k\ran$ and
$|k^\prime\ran$ orthonormal we have
\begin{eqnarray}
\| \ket{k_p} \|^2 = |\bra{k} k_p\ran|^2 + \sum_{k^\prime} 
|\bra{k^\prime} k_p\ran|^2 = p \;, \label{NUOVA}
\end{eqnarray}
where the sum is performed over the set  of vectors~$\{|k^\prime\ran\}_{k^\prime}$ which
complete $|k\ran$ to an orthonormal basis of ${\cal H}_S$. 
It should be stressed that the above derivation holds for all value of $p<1$ and  for 
Hilbert space  ${\cal H}_S$ (also infinite-dimensional Hilbert spaces).

\section{Mixed outputs}\label{s:due}

Consider now the case of   generalized 
quantum movers $\mbox{GQM}(p)$ with not necessarily pure
output states $\rho_b$.
These mappings include all those transformations which implement a proper quantum mover
transformation~(\ref{THETRA}) 
probabilistically, i.e. where the output state is randomly selected among the set
of the vectors $|\psi_p\ran$ which satisfies the condition of Eq.~(\ref{unnot}).  
Differently from the previous case, these generalized quantum 
movers are 
allowed at least for some restricted values of $p$. 
 What is more interesting 
is that the generalized quantum movers $\mbox{GQM}(p)$ can be implemented by means of 
LCPT maps only if $p$ exceeds a threshold $p_{\mbox{\small t}}(d)$ which  explicitly 
depends on $d$. 
Here we provide both upper and 
lower bounds for this threshold, i.e.
\begin{eqnarray}
 1/(2 d-1)\; \leqslant \; p_{\mbox{\small t}}(d) \;\leqslant \; 1/(d +1)  \label{UPLOW}\;.
\end{eqnarray}
Such bounds coincide for $d=2$ so that   $p_{\mbox{\small t}}(2)=1/3$
and the threshold of the qubit case is completely determined by the above inequalities.
As in the case of Eq.~(\ref{ULTIMA}) this is 
a direct consequence of the $2/3$ bound on  the optimal fidelity associated with 
an approximate realization of a U-NOT~\cite{BUZEK,PASQ}.

To derive the upper bound of Eq.~(\ref{UPLOW}) it is sufficient to provide a  LCPT 
transformation that satisfies Eq.~(\ref{movers1}) for $p\geqslant 1/(d+1)$. 
For instance let us analyze the  linear mapping which takes any trace-class 
operator $\Lambda$ of ${\cal H}_S$ to
\begin{eqnarray}
	{\cal N}(\Lambda) \equiv \frac{dp -1}{d-1} \Lambda + 
	\frac{1-p}{d-1} \mbox{Tr}[ \Lambda ] \; \openone \;,
\label{esempio}
\end{eqnarray}
where $\openone$ is the identity operator of ${\cal H}_S$. The transformation ${\cal N}$ 
belongs to the set of universal inverter super-operators that have been introduced in 
Refs.~\cite{BUZEK1,HORO} in the context of entanglement criteria for bipartite systems.
Moreover, for $p=1/(d+1)$ Eq.~(\ref{esempio})  gives the approximated U-NOT 
mapping~\protect\cite{BUZEK,PASQ,DEMA} which provides optimal output fidelity for $d=2$ 
and, in the case $d>2$, provides optimal fidelity in the 
class of  LCPT maps  which are covariant   
with respect to arbitrary unitary transformations of ${\cal H}_S$~\cite{BUZEK1}.
One easily verifies that the transformation (\ref{esempio}) 
satisfies the condition~(\ref{movers1}) 
for all~$|\psi\ran$. However, it turns out~\cite{BUZEK1} that the map ${\cal N}$ 
is completely positive only for 
$p\geqslant \; 1/(d +1)$ and this fixes the upper bound for the threshold 
$p_{\mbox{\small t}}(d)$.
It should be stressed that in agreement with the no-go theorem 
on the deterministic quantum movers, the map ${\cal N}$ 
of Eq.~(\ref{esempio}) transforms pure states of ${\cal H}_S$ in mixed outputs.
In fact, the purity of the output state is
$$\mbox{Tr} \left[ \; 
{\cal N}(|\psi\ran\lan\psi|)^2 \; \right] = \left(\frac{1-p}{d-1}\right)^2 +
p^2, $$
which is strictly less than $1$ for  $p\geqslant \; 1/(d +1)$  and $d\geqslant 2$.

To derive the lower bound of Eq.~(\ref{UPLOW})
 assume that a LCPT map $\cale$ satisfying Eq.~(\ref{movers1}) for all 
input states $|\psi\ran$ exists.
By applying this relation to the vector of Eq.~(\ref{VETTORI}) we get the
following equation
\begin{eqnarray}
p&=&|\alpha|^2 \bra{k}\: \cale (|\psi\ran\lan\psi|)\; |k\ran
+ |\beta|^2 \bra{k^\prime}\: \cale (|\psi\ran\lan\psi|)\; |k^\prime\ran \nonumber \\
&&+\; [\; \alpha \beta^* \bra{k^\prime}\: \cale (|\psi\ran\lan\psi|)\; |k\ran + \mbox{c.c.} \; ]\;,
\label{equazione}
\end{eqnarray} 
with
\begin{eqnarray} 
\cale (|\psi\ran\lan\psi|) &\equiv& |\alpha|^2 \cale (|k\ran\lan k|) 
+  |\beta|^2 \cale (|k^\prime \ran\lan k^\prime |)  \nonumber\\
&&+\; 
 [ \; \alpha \beta^*  \cale (|k\ran\lan k^\prime|) + \mbox{h.c.} \; ] \;.\label{equazione1}
\end{eqnarray}
Again by requiring Eq.~(\ref{equazione}) to hold for all $\alpha$ and $\beta$ we get the following 
constraints on the map $\cale$ which apply for all $|k\ran$ and $|k^\prime\ran$
orthonormal 
\begin{eqnarray}
\left\{ 
\begin{array}{l}
2p = \lan k k^\prime k^\prime k\ran
+ \lan k^\prime k k k^\prime \ran
+ (\; \lan k k k^\prime k^\prime \ran + c.c \;)\\
\lan k k^\prime k k^\prime \ran= 0 \\
\lan k k k k^\prime \ran +\lan k k^\prime k k\ran 
 =0
 \;,
\end{array}\right. \label{equo}
\end{eqnarray}
where we defined the quantities
$\lan j k \ell m \ran \equiv \bra{j} \cale (|k \ran\lan \ell|)  |m\ran 
= \lan m \ell k j \ran^*$ (see Appendix~\ref{appendixB}).
The previous equations can be supplemented by another constraint which can be obtained 
by applying  $(\cale_S\otimes{\cal I}_A)$ 
to the entangled state 
$$|\Psi\ran_{SA} \equiv \frac{1}{\sqrt{d}}\sum_{k=1}^d |k\ran_S\otimes |k\ran_A,$$
and requiring its trace 
to be properly normalized to 1
[In the above expression $\{|k\ran_{S,A}\}_k$ are orthonormal bases of 
${\cal H}_S$ and ${\cal H}_A$,
respectively, and ${\cal I}_A$ is the identity mapping on the ancillary
space ${\cal H}_A$].
 This yields
\begin{equation}
1= \frac{1}{d} \sum_{k=1}^{d}\sum_{k^\prime=1}^d \lan k^\prime kk k^\prime \ran 
 =  p + \sum_{k \neq k^\prime} 
\lan k^\prime kk k^\prime \ran /d \;,
\label{traccia} 
\end{equation} 
where we used the hypothesis~(\ref{movers1}) to write 
$\lan k kk k \ran = p$ for all $|k\ran$.
Equation~(\ref{traccia}) can be further simplified by replacing the first relation
of the system~(\ref{equo}). This gives 
$$ \sum_{k \neq k^\prime} 
\lan k k k^\prime k^\prime \ran = d(d p -1).$$
The lower bound of Eq.~(\ref{UPLOW}) finally follows by comparing this
relation with the inequality
$
|\lan k k k^\prime k^\prime\ran | \leqslant p
$, obtained  by introducing a Kraus decomposition
of the LCPT map $\cale$ 
and applying the Cauchy-Schwarz inequality to the left hand side term~\cite{FOOTNOTE4}. 

We finally comment on the relations of  $\mbox{GQM}(p)$ with the U-NOT. One can easily verify 
that the set of the linear transformations which implement $\mbox{GQM}(p)$
is closed under convex combination, e.g. given ${\cal E}_{p_i}$ LCPT implementations 
of $\mbox{GQM}(p_i)$ on S and given $\{\lambda_i\}_i$ a set of probabilities,
the transformation  $\sum_i \lambda_i {\cal E}_{p_i}$ 
is a LCPT map which
implements  the generalized mover  $\mbox{GQM}(p)$ on S 
with $p=\sum_i \lambda_i p_i$ being the
average of the $p_i$'s.
This observation allows us to establish that if the U-NOT could be implemented through 
a LCPT transformation, then convex combinations of such a map and the identity map will 
implement the $\mbox{GQM}(p)$ for $p<1$.

\section{Quantum Movers and Entanglement}\label{s:tre}
In this last section we discuss an application of quantum movers in the
context of entanglement witness. 
A central issue in quantum information 
is to determine whether or not a given density matrix $R$ of a 
composite system S+A is entangled. The identification of classes of impossible 
transformations 
 can become a 
powerful tool in this respect~\cite{HORO3}. In our case, 
one can construct a one parameter set of positive 
{\em but} not completely positive linear maps $\{{\cal M}^{(p)}\}_{p}$ which 
implement the generalized quantum mover transformation $\mbox{GQM}(p)$ on the system S.
The presence of entanglement in $R$ will then be detected by 
looking for the presence of negative eigenvalues in  ${\cal M}_S^{(p)}\otimes {\cal I}_A(R)$ 
for some values of $p$. 
Moreover, one can use $p$  as an effective ``measure'' of the entanglement in $R$ as one 
expects intuitively only highly entangled states to feel the non complete positivity of 
the transformation ${\cal M}^{(p)}$ with $p\sim 1$. A suitable choice for the operators 
$\{{\cal M}^{(p)}\}_{p}$ is obtained by adding to the transformations~(\ref{esempio}) 
a linear term which gives 
null contribution to Eq.~(\ref{movers1}), e.g.
\begin{eqnarray}
{\cal M}^{(p)} (\Lambda) \equiv \frac{dp -1}{d-1} \Lambda + 
\frac{1-p}{d-1} \mbox{Tr}[ \Lambda ]\; \openone + i [ \Lambda, \Theta ]
\label{NEWesempio1}
\end{eqnarray}
with $\Theta$ being a Hermitian operator on ${\cal H}_S$. 
For the sake of simplicity let us consider the $d=2$ case where we can select 
$\Theta = \sqrt{p(1-p)} \; \sigma_{\hat{n}}$ with $\sigma_{\hat{n}}$ being the Pauli 
operator along the $\hat{n}$ direction. With this choice Eq.~(\ref{NEWesempio1}) yields a  
mapping which, for all $p\in[0,1[$ is positive but not completely positive (this can be 
verified for instance by a direct inspection of the corresponding Choi 
matrix~\cite{CHOI}). As a test consider the set of Werner states~\cite{WER}
\begin{eqnarray}
R_q \equiv  \frac{1-q}{4} \, \openone_S \otimes \openone_A 
\, + \, q \, |\Psi_-\rangle_{SA} \langle \Psi_-| \; 
\label{werner},
\end{eqnarray}
which  have concurrence~\cite{CONC} $C= \max \{ 0, (3q-1)/2\}$ and are hence 
separable if and only if $q\leqslant 1/3$  
[in the above expression $|\Psi_-\rangle = (|01\rangle_{SA} - |10\rangle_{SA})/\sqrt{2}$
is the singlet state].
The eigenvalues of  ${\cal M}_S^{(p)}\otimes {\cal I}_A(R_q)$ can
be easily computed: three of them are always positive and the fourth
is equal to
\begin{eqnarray}
\lambda(q,p)\equiv  q(p-(3q-1)/2q)/2\;. 
\end{eqnarray}
For $R_q$ separable (i.e. $q\leqslant 1/3$) this quantity is positive for all $p$. 
However for  $R_q$ entangled (i.e. $q> 1/3$) $\lambda(q,p)$ is negative for all 
$p$ smaller than the critical value $(3 q -1)/2q=(3/2) [ 1-1/(2 C+1)]$ which
monotonically increases with $C$. As intuitively expected the more entanglement is  in $R_q$
the more transformations ${\cal M}_S^{(p)}\otimes {\cal I}_A(R_q)$ 
are able to detect it. 

\section{Conclusions}\label{s:conc}
We have introduced a new class of impossible
quantum operations, the quantum movers which include, as a   
special case, the U-NOT transformation of Refs.~\cite{BUZEK,PASQ}.
We proved that it is not possible to transform an unknown state $|\psi\ran$
of a system S under the requirement that the final output configuration $|\psi_p\ran$
has a fixed (non unitary) component along the direction of the initial state.
By relaxing the requirement on the purity of the final configuration of the system,
there exists a threshold $p_{\mbox{\small t}}(d)$ for the value of the input-output fidelity 
above which mixed quantum movers can be implemented. We related the movers to the U-NOT
and we showed an application of these transformation in the context of separability
criteria.
The approach introduced in this work can be further extended. One can, for example, 
analyze the case in which the output fidelity of a quantum mover is bounded 
in a given interval, e.g. $F\in [p_1,p_2]$. Moreover, as in the case of a U-NOT
transformation, 
it may be interesting to analyze under which conditions the $\mbox{GQM}(p)$ with $p$ 
below the threshold value $p_t(d)$ can be realized approximatively.

\appendix
\section{}\label{appendixA}
Here we show that if Eq.~(\ref{cpt10})  applies for all the complex amplitudes 
$\alpha$ and $\beta$
then the constraint~(\ref{cpt11}) must be satisfied. 
For ease of notation define $\eta_1 \equiv  \bra{k^\prime} k_p\ran$, 
$\eta_2 \equiv \bra{k} {k^\prime}_p\ran$ and
$h \equiv {_A\!\bra{\phi(k^\prime)}}\phi(k)\ran_A$ and call $F(\alpha, \beta)$
the function at the right hand side of Eq.~(\ref{cpt10}).
According to this equation $F(\alpha, \beta)$ is equal to $p$ for all values 
of $\alpha$ and $\beta$.
Consequently we can write
\begin{eqnarray} 
\left\{ 
\begin{array}{l}
 F(\alpha,\beta)-F(\alpha,-\beta)=0  \\
 F(\alpha,\beta)+ F(\alpha,-\beta) -  F(\alpha,i\beta)-F(\alpha,-i\beta)=0 
\;. 
\end{array}
\right.
\label{appendix1}
\end{eqnarray}
When considered for generic $\alpha, \beta$ and for $p\neq 0$ the above system reduces
to
\begin{eqnarray} 
\left\{ 
\begin{array}{l}
\eta_1 \;\eta_2^* \; h =0 \\
\eta_1 + \eta_2^*\; h = 0 \\
\eta_2 + \eta_1^* \;h = 0 \;,
\end{array}
\right. 
\nonumber 
\end{eqnarray}
which has solution $\eta_1=\eta_2=0$ independently from the value of $h$. 
This proves that Eq.~(\ref{cpt10}) implies Eq.~(\ref{cpt11}) for $p\neq 0$.
For $p=0$  we replace Eq.~(\ref{appendix1})
with
\begin{eqnarray} 
F(\alpha,\beta)+F(\alpha, i \beta)=4 |\alpha \beta|^2 ( |\eta_1|^2 + |\eta_2|^2 ) = 0 \nonumber\;.
\end{eqnarray}
which, again, holds for all $\alpha$ and $\beta$ if and only if $\eta_1=\eta_2=0$.
 
\section{}\label{appendixB} Here we show that if Eq.~(\ref{equazione})  
applies for all the complex amplitudes 
$\alpha$ and $\beta$,
then the constraints~(\ref{equo}) must be satisfied.
Consider the right hand side term of Eq.~(\ref{equazione}). By using Eq.~(\ref{equazione1}),
this can be explicitly written as,
\begin{eqnarray} 
G(\alpha,\beta) &\equiv & (|\alpha|^4+|\beta|^4) p/2 
\nonumber \\
&+ &
|\alpha^3 \beta| (e^{i\theta} \lan kkkk^\prime\ran + e^{-i \theta} \lan kkk^\prime k\ran)
\nonumber \\
&+ &
|\alpha \beta^3| ( e^{-i\theta} \lan k^\prime k^\prime 
k^\prime k \ran + e^{i \theta} 
\lan k^\prime k^\prime k k^\prime\ran)
\nonumber \\
&+ & 
|\alpha \beta|^2 ( \lan k k^\prime 
k^\prime k \ran + \lan kkk^\prime k^\prime \ran + e^{2i \theta} 
\lan k k^\prime k k^\prime \ran) 
\nonumber \\
&+ & \mbox{c.c.}
\label{funzioneG}
\end{eqnarray}
where $\theta\equiv \arg (\alpha^* \beta)$ is the relative phase of $\beta$ and $\alpha$.
According to Eq.~(\ref{equazione}) the function $G(\alpha,\beta)$ is constant and equal to
$p$ for all values of the amplitudes $\alpha$ and $\beta$.
The constraints of Eq.~(\ref{equo}) then follows by studying the 
Fourier components of~(\ref{equazione}) with respect to the phases $0,\pm \theta$ and  $\pm 2 \theta$,
or equivalently by considering the following identities
\begin{eqnarray} 
&&G(\alpha,\beta)+ G(\alpha,-\beta) + G(\alpha,i\beta) + G(\alpha,-i\beta)= 4p 
\nonumber \\
&&G(\alpha,\beta)+ G(\alpha,-\beta) -  G(\alpha,i\beta)-G(\alpha,-i\beta)= 0  
\nonumber \\
&&G(\alpha,\beta)- G(\alpha,-\beta)=0\;.\label{identity1}
\nonumber
\end{eqnarray}

\acknowledgments
This work was supported by the European Community under contracts,
IST-SQUBIT2 and RTNANO.


\begin{thebibliography}{99}
\bibitem{WERNERBOOK}R. F. Werner, {\em Quantum Information Theory -- an Invitation} 
	in Springer Tracts in Modern
	Physics Vol.  173, (Springer-Verlag, Telos, 2001), p. 14.
\bibitem{WOOTTERS}W. K. Wootters and W. H. Zurek,
	Nature {\bf 299}, 802 (1982).
\bibitem{PASQ}
	N. Gisin and S. Popescu, Phys. Rev. Lett. {\bf 83}, 432 (1999);
	H. Bechmann-Pasquinucci and N. Gisin, Phys. Rev. A {\bf 59}, 4238 (1999).
\bibitem{BUZEK}
	V. Bu\v{z}ek, M. Hillery, and R. F. Werner, Phys. Rev. A {\bf 60}, R2626 (1999);
	V. Bu\v{z}ek, M. Hillery, and R. F. Werner, J. Mod. Opt. {\bf 47}, 211 (2000).
\bibitem{NOTAQRP}For instance the fidelity $F$  among the input 
	$\rho_{in}$ and the
	output state $\rho_{out}$ of the transformation should fulfill predefined constraints, e.g.
	$p_1\leqslant F \leqslant p_2$ with the parameters $p_1$ and $p_2$ independent from $\rho_{in}$.
\bibitem{HORO3}M. Horodecki, P. Horodecki,  and R. Horodecki, Phys. Lett. A {\bf 223}, 1 (1996).
\bibitem{HORO}
	M. Horodecki and P. Horodecki, Phys. Rev. A {\bf 59}, 4206 (1999). 
\bibitem{BUZEK1}
	P. Rungta, V. Bu\v{z}ek, C. M. Caves, M. Hillery, and G. J. Milburn, 
	Phys. Rev. A {\bf 64}, 042315 (2001).
\bibitem{JOZSA}
	R. Jozsa, J. Mod. Opt. {\bf 41}, 2315 (1994); 
	A. Uhlmann, Rep. Math. Phys. {\bf 9}, 273 (1976).
\bibitem{CHUANG}M. A. Nielsen and I. L. Chuang, {\em Quantum
	Computation and Quantum Information} 
	(Cambridge University Press, Cambridge, 2000).
\bibitem{DEMA}
	F. De Martini, V. Bu\v{z}ek, F. Sciarrino, and C. Sias, Nature {\bf 419}, 815 (2002).
\bibitem{NOTA0}
	The trace decreasing operation of ${\cal H}_S$ describes non-deterministic
	processes which are excluded from the analysis by hypothesis.
\bibitem{FOOTNOTE4}Let $\{M_\ell\}_\ell$ be a Kraus set of $\cal E$, such that
	${\cal E}(\Lambda)= \sum_{\ell} M_{\ell} \Lambda M_{\ell}^\dag$~\cite{CHUANG}. 
        Then using the  Cauchy-Schwarz inequality we have
	\begin{eqnarray}	
	|\lan k k k^\prime k^\prime \ran| 
	&=& |\sum_\ell  \lan k |M_\ell |k \ran \lan k^\prime|  M_\ell^\dag |k^\prime\ran| 
	\nonumber \\
	&\leqslant& \sqrt{ \sum_\ell   |\lan k |M_\ell |k \ran|^2 \; \sum_\ell   |\lan k^\prime 
	|M_\ell |k^\prime \ran|^2 }  \nonumber \\
	&=&  \sqrt{\langle k k k k \rangle  
	\langle k^\prime k^\prime k^\prime k^\prime \rangle } \nonumber 
	= p\;. \end{eqnarray} 
\bibitem{CHOI}M. D. Choi, Lin. Alg. Appl. {\bf 10}, 285 (1975).
\bibitem{WER}R. F. Werner, Phys. Rev. A {\bf 40}, 4277 (1989).
\bibitem{CONC}W. K. Wootters, Phys. Rev. Lett. {\bf 80} 2245 (1998).
\end{thebibliography}
\end{document}